\documentclass[aps,prl,twocolumn,longbibliography]{revtex4-1}

\usepackage{natbib}
\usepackage{xspace}

\usepackage[sumlimits]{amsmath}
\usepackage{amsfonts,amssymb,amsthm}
\usepackage{graphicx}
\usepackage[margin=1in,footskip=0.25in]{geometry}

\usepackage{xr}

\newcommand{\TJWat}{IBM T.J. Watson Research Center, Yorktown Heights, NY 10598, USA}

\newcommand{\calP}{{\mathcal P}}
\newcommand{\calC}{{\mathcal C}}
\newcommand{\calW}{{\mathcal W}}
\newcommand{\calw}{{\mathcal Y}}
\newcommand{\PP}{P}
\newcommand{\ZZ}{$\mathbf{ZZ}$\xspace}
\newcommand{\srb}{simRB\xspace}
\newcommand{\crb}{corrRB\xspace}
\newcommand{\PTM}{PTM\xspace}

\newcommand{\figfolder}[1]{}

\begin{document}

\title{Correlated Randomized Benchmarking}

\author{David C. McKay}
\email{dcmckay@us.ibm.com}
\author{Andrew W. Cross}
\author{Christopher J. Wood}
\author{Jay M. Gambetta}
\affiliation{\TJWat}
\date{\today}

\begin{abstract}
To improve the performance of multi-qubit algorithms on quantum devices it is critical to have methods for characterizing non-local quantum errors such as crosstalk. To address this issue, we propose and test an extension to the analysis of simultaneous randomized benchmarking data --- correlated randomized benchmarking. We fit the decay of correlated polarizations to a composition of fixed-weight depolarizing maps to characterize the locality and weight of crosstalk errors. From these errors we introduce a crosstalk metric which indicates the distance to the closest map with only local errors. We demonstrate this technique experimentally with a four-qubit superconducting device and utilize correlated RB to validate crosstalk reduction when we implement an echo sequence. 
\end{abstract}

\maketitle

In order to build effective quantum computers it is necessary to detect and correct errors in noisy hardware. Error correcting codes, such as the surface code\cite{dennis:2002,fowler:2012}, have been devised to protect against noise provided the error rates on the underlying physical qubits are below the threshold of the code. However, what are the appropriate error rates to measure? For typical hardware realizations that have a basis set of one- and two-qubit gates, e.g., superconducting qubits, it is standard to only measure the errors on the gates with this set. While this is efficient, it does not characterize non-local noise and is typically not sufficient for estimating the overall performance of multi-qubit circuits\cite{mckay:2019}. Non-local errors may reduce the effectiveness of quantum codes and/or increase the overhead of using those codes. For example, the surface code threshold for nearest-neighbor depolarizing noise is lower than the threshold for local depolarizing noise, even when the decoding procedure is optimized for each model\cite{maskara:2018,tuckett:2018,debroy:2019}. \\

For a quantum system, errors can be coherent, described by extra terms in the Hamiltonian, or incoherent, described by dissipators in a master equation. For multi-qubit systems, the operators for these Hamiltonian error terms can be written as tensor products of Pauli operators $\sigma_0^{i_0} \otimes \sigma_1^{i_1} \otimes \ldots \sigma_n^{i_n}$ where $i=0,1,2,3$ are the $\mathbf{I}$, $\mathbf{X}$, $\mathbf{Y}$ and $\mathbf{Z}$ operators. These error terms operate on subsets of qubits, and are classified by weight and locality. The weight is the the number of non-identity Pauli operators in the subset. For example, a weight 1 crosstalk error is a term such as $\sigma_{0}^{0} \otimes \sigma_{1}^{3} = \mathbf{I} \mathbf{Z}$, which could occur due to Stark shifts on a qubit from the drive tone on another qubit. Weight 2 crosstalk errors are due to qubit-qubit interactions that are not completely turned off between intentional two-qubit gates, e.g., an always on \ZZ interaction. Weight 2 terms lead to unwanted entanglement which is difficult to correct. Geometric locality is the average number of edges on the graph between each pair of qubits in the subset. For example, a weight 2 term on a linearly connected three-qubit device could be $\mathbf{I} \mathbf{Z}\mathbf{Z}$ (1 edge) versus $\mathbf{Z} \mathbf{I} \mathbf{Z}$ (2 edges). When detecting crosstalk it is important to measure both the weight and locality of the error terms. In general, full characterization of crosstalk requires n-qubit process tomography which is impractical for large $n$ since the map has an exponential number of parameters. Furthermore, there are also an exponential number of different process maps to consider because each gate (and combination of gates in parallel) potentially has a distinct error map. For example, if each qubit is either idle ($\mathbf{I}$) or performing the same single qubit unitary $\mathbf{U}$, there are $2^n$ maps to measure. Recent work has focused on detecting the presence of crosstalk between subsystems in large devices \cite{sarovar:2019,rudinger:2019}, however these methodologies lack quantitative estimates for the magnitude of the crosstalk error terms. \\

Here we consider a method to measure the average crosstalk errors over the Clifford gateset and classify these errors in terms of weight and locality via simultaneous randomized benchmarking (\srb)~\cite{magesan:2011,gambetta:2012}. In \srb, independent RB sequences are run simultaneously on $m$ subsystems of an $n$ qubit system ($1\le m \le n$), and the decay of polarization in each subsystem is fit to an exponential decay versus sequence length to obtain $m$ fidelities ($F_{m,s}$). Alternatively, we can run RB in the same $m$ subsystems when the other $m-1$ subsystems are idle (in $m$ separate experiments) and obtain fidelities $F_{m,i}$. The standard procedure proposed in Ref.~\cite{gambetta:2012} is to compare $F_{m,s}$ to $F_{m,i}$ and if these quantities are different then this indicates crosstalk errors, however, this does not give a rigorous metric for crosstalk. As discussed in Ref.~\cite{gambetta:2012}, when the other subsystems are idle the full $n$ qubit map is not properly twirled.   \\

Here we propose and test an extension to \srb --- correlated RB (\crb). In \crb we perform $m$ subsystem \srb and fit the decay of the correlated polarizations (e.g. for a 3-qubit system $\langle ZIZ\rangle$)  to get all $2^m$ independent decay parameters of the twirled map. Fitting these parameters to a specific model gives the crosstalk errors. The paper is organized as follows. First, we describe the theory of \crb, introduce a model error map for classifying \crb results and propose a crosstalk metric. Next, we implement the \crb procedure experimentally for single qubit gates on a four-qubit superconducting circuit using fixed-frequency transmons. From these measurements we show that the dominant crosstalk errors are weight 2, consistent with the measured \ZZ terms in the Hamiltonian. We test that \crb can measure high weight errors by purposely injecting weight 4 noise and successfully observing these in our error model after fitting the data. Finally, we implement an echo sequence designed to cancel weight 2 errors from the \ZZ interaction and use \crb to verify its success. Overall, correlated RB appears to be a valuable tool for measuring and validating the reduction of crosstalk errors in multiqubit devices, which will be critical for executing large algorithms on quantum devices. In preparation of this manuscript we were made aware of similar work performed in Ref.~\cite{harper:2019}. 

\section{Theory}

The challenge of quantum characterization is to estimate parameters of a general quantum error process which can be used to assess the overall performance of a quantum computation. Full characterization involves reconstructing the complete description of the error map though a procedure such as quantum process tomography, which is impractical more than a handful of qubits, hence it is important to consider partial characterization methods which instead directly estimate useful performance metrics which may be derived from the full process. One such metric is the average gate fidelity of the operation\cite{magesan:2012},
\begin{equation}
	F = \int \mathrm{Tr}\left(|\phi\rangle \langle \phi| \Lambda[|\phi\rangle \langle \phi|] \right)d\mu(\phi)
\end{equation}
which quantifies how close the physical error map is to the identity. Although this is a useful overall assessment of the system performance, to diagnose errors we need more fine-grained characterization tools. In particular, for multi-qubit systems we are interested in measuring crosstalk, which has important implications for error correction codes. Crosstalk errors are those errors that occur outside the subsystem of the intended gates. For a system partitioned into $m$ subsystems we propose a crosstalk metric,
\begin{equation}
	\eta(\Lambda) := \underset{\Gamma_j}{\mathrm{min}}\lVert\Lambda-\bigotimes_{j}\Gamma_j\rVert_\diamond \label{eqn:eta}
\end{equation}
where each $\Gamma_j$ is a channel contained in subsystem $j$. The metric measures how far a channel $\Lambda$ is from the nearest tensor product channel. 

For a general process, measuring the map and calculating $\eta$ is intractable since even describing $\Lambda$ is intractable. Fortunately, simultaneous randomized benchmarking (\srb)~\cite{gambetta:2012}, which is an extension of standard RB~\cite{magesan:2011}, transforms the map via the twirl operator into a convenient form for dissecting crosstalk errors. To discuss \srb we introduce the notation 
\begin{equation}
B_m = \{\{i,j,\ldots\},\ldots,\{k,l,\ldots\}\}
\end{equation}
which is a partition of the $n$-qubit system into $m$ subsystems. The number of qubits in the $i^{\mathrm{th}}$ subsystem is $n_i$. For example to denote 1Q \srb on a 3-qubit system we would write the partition as $\{\{0\},\{1\},\{2\}\}$. The procedure to run \srb on a n-qubit system partitioned into $m$ subsystems is as follows. For the $i^{\mathrm{th}}$ subsystem, create a sequence of $l-1$ randomly selected gates from the $n_i$-qubit Clifford group ($\{C_{i,j}\}$); calculate the inverse of this sequence $C_{i}^{-1}$ and append it to the end of the sequence so that the full set of gates is ideally the identity operator. Starting with the qubits in the ground state apply each subsystem sequence simultaneously and measure the polarization of the qubits in each subsystem (e.g. $\langle ZIII\ldots \rangle$). Repeat for different random sequences (trials) and different sequence lengths $l$ and then average the results across these trials. Fit the average polarization in subsystem $i$ versus $l$ to the curve $A\alpha_i^{l}+B$ to get the decay parameter $\alpha_i$, which is related to the average gate error in that subsystem as $(2^{n_i}-1)/2^{n_i} (1-\alpha_i)$. From this procedure we get $m$ decay parameters. As shown in Ref.~\cite{gambetta:2012}, the twirled map obtained by performing this \srb procedure is a Pauli channel,
\begin{equation}
	\tilde{\Lambda}(\rho) = \frac{1}{|\calC_{n_i}^{\otimes m}|}\sum_{U\in\calC_{n_i}^{\otimes m}}U^\dagger\Lambda(U\rho U^\dagger)U = \sum_{S\subseteq B_m} p_S\calW_S(\rho), \label{eqn:twirledmap}
\end{equation}
on subsystems $S\subseteq B_m$, i.e., $S$ is a tensor product of subsystems in $B_m$. Since there are $2^m$ subsystems $S$ with coefficients $p_S$, this means there is some information from Eqn~(\ref{eqn:twirledmap}) that is not measured in the standard \srb procedure. The goal of \crb is to measure these extra coefficients.  The weight of $S$ ($|S|$) is the number of subsystems from $B_m$ in $S$. The locality is the average number of edges on a graph connecting each pair of subsystems. In~(\ref{eqn:twirledmap}),  $\calW_S$ is the fixed-subspace-weight channel (the sum of all Pauli channels in $S$),
\begin{eqnarray}
	\calW_S(\rho) & = & \frac{1}{\prod_{\{i|i\in S\}} (4^{n_i}-1)} \calw_S(\rho), \label{eqn:fixedweight} \\
	\calw_S(\rho) & = & \sum_{\substack{\PP\in\calP_n\\\mathrm{supp}(\sigma)=S}}\PP\rho\PP^\dagger.
\end{eqnarray}
If $S=\emptyset$, define $\calW_\emptyset(\rho)=\rho$. The map in (\ref{eqn:twirledmap}) is completely positive and trace preserving (CPTP) if each $p_S\in [0,1]$ and $\sum_{S\subseteq B_m} p_S=1$. Instead of set notation we can also write $S$ in bitstring notation where the length of the string is equal to $m$ and the bit is $1$ if that subsystem is part of the space $S$; the weight of $S$ is then the number of 1's in the string. For example, if the partition is $\{\{0\},\{1\}\}$ then $S=00,10,01,11$ and the twirled map after \srb is,
\begin{eqnarray}
	\tilde{\Lambda}(\rho) & = & p_{00} \rho + p_{10} \frac{\mathbf{XI} \rho \mathbf{XI} + \ldots}{3} + \ldots \nonumber \\
 & & p_{01} \frac{\mathbf{IX} \rho \mathbf{IX} + \ldots}{3} + p_{11} \frac{\mathbf{XX} \rho \mathbf{XX} + \ldots}{9}.
\end{eqnarray}
\\

Experimentally, we measure the $\mathbf{Z}$ correlators for subsystem $S$ after applying the map (\ref{eqn:twirledmap}) $l$ times starting from an initial density matrix $\rho_0$. Composing maps is multiplicative in the the Pauli transfer matrix (\PTM) representation. The \PTM representation of a channel $\Lambda$ is 
\begin{equation}
(R_\Lambda)_{i,j}=\frac{1}{2^n}\mathrm{Tr}(\PP_i\Lambda(\PP_j))
\end{equation}
where $\PP\in\calP_n$ (the n-qubit Pauli's). If $\mathbf{P}_S=\sum_{i \in S} |i \rangle \langle i|$ is a projector onto the same subspace as our fixed-weight channel (\ref{eqn:fixedweight}) we can write the twirled channel (\ref{eqn:twirledmap}) as a diagonal matrix in the \PTM as
\begin{equation}
	R_{\tilde{\Lambda}} = \sum_{S\subseteq B_m} \alpha_S \mathbf{P}_S, \label{eqn:ptmmap}
\end{equation}
and applying the map $l$ times gives,
\begin{equation}
	R_{\tilde{\Lambda}^l} = \sum_{S\subseteq B_m} \alpha_S^l \mathbf{P}_S. \label{eqn:ptmmap_l}
\end{equation}
Noting that $\alpha_S$ is related to the untwirled map $\Lambda$ as $\alpha_S=\mathrm{Tr}(\mathbf{P}_S R_\Lambda)/\mathrm{Tr}(\mathbf{P}_S)$. In the \PTM representation the final density matrix is a vector $r_{\rho_f}=R_{\Lambda} \cdot r_{\rho_0}$ where $r_{\rho}$ is a vector of the Pauli coefficients of $\rho$. Then the correlator of the $i^{\mathrm{th}}$ Pauli $\langle P_i\rangle$ is $r_{\rho, i}$ and so the output of a \srb experiment is,
\begin{equation}
\langle \mathbf{Z}_i \rangle = \alpha_S^l \gamma_i
\end{equation}
where $\mathbf{Z}_i$ refers to one of the $Z$ Pauli operators which is in subsystem $S$ and $\gamma_i$ is the projection of $\rho_0$ onto $\mathbf{Z}_i$ ($\gamma_i=1$ if $\rho_0 = |0\rangle \langle0|$). Note that the measurement of each correlator will be influenced by measurement noise.\\

\begin{figure}
\includegraphics[width=0.45\textwidth]{\figfolder{1a}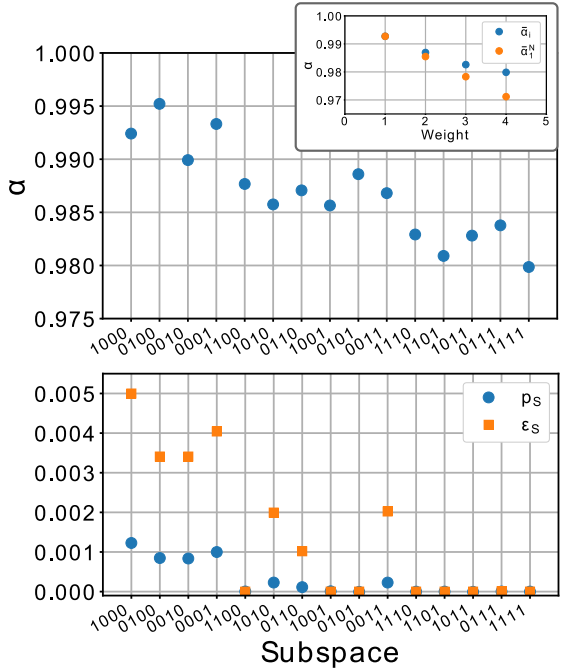}
\caption{(Color Online) Different representations of the twirled map from simulated $\{\{0\},\{1\},\{2\},\{3\}\}$ 4Q correlated RB data, i.e., 1Q SimRB on all qubits. The data is simulated using correlated coherent noise (\ZZ) and incoherent uncorrelated noise ($T_1$ and $T_2$) to match the experimental values; see the experiment section for more details. (Top) The $\{\alpha\}$ parameters measured directly from fitting the correlated SimRB data. Individually these do not reveal obvious crosstalk information. (Top Inset) If the noise is completely local then the geometric mean of the weight $n$ $\alpha$ parameter will be $\bar{\alpha}_1^n$. Since the data deviates from this trend, crosstalk exists but is difficult to localize and quantify. (Bottom) Pauli fixed-weight representation $p_S$ and the $\epsilon_S$ depolarizing fixed-weight representation. These representations clarify the locality and weight of the crosstalk errors. The $\epsilon$ representation completely removes local terms from the higher weight elements.  For this data we calculate the Pauli crosstalk metric (\ref{eqn:eta2}) and get a value of $\approx 0.009$. \label{fig:simdata}}
\end{figure}

The idea behind correlated RB (\crb) is to measure the higher weight decay parameters (e.g. $\alpha_{11}$) and transform them to coefficients of an error model that is easy to relate to the crosstalk metric. We start by relating the parameters $\{\alpha_S\}$ of the PTM to the parameters $\{p_S\}$ of the Kraus form. The \PTM representation of a Pauli channel $\Lambda_P = \sum_{i} p_i \PP_i \rho \PP_i$ is,
\begin{equation}
(R_{\Lambda_P})_{j,j}= \sum_{i} p_i (-1)^{\omega(\PP_j,\PP_i)} 
\end{equation}
which is diagonal and where $\omega(\PP_j,\PP_i)=0$ if $[\PP_j,\PP_i]=0$ and $\omega(\PP_j,\PP_i)=1$ otherwise. This follows from the definition of the \PTM. To go back to the Kraus representation,
\begin{eqnarray}
	p_{i} & = & \frac{1}{4^n}\mathrm{Tr}(R_{\PP_{i}\rho\PP_i}R_{\Lambda_P}) \nonumber \\
       & = & \frac{1}{4^n}\sum_{j=1}^{4^n} (-1)^{\omega(\PP_i,\PP_j)} (R_{\Lambda_P})_{j,j}.
\end{eqnarray}
Sample simulation data of the different values of $\alpha_S$ and the corresponding $p_S$ are given in Fig.~\ref{fig:simdata} for a 4-qubit system.\\

The issue with the $\alpha_S$ ($p_S$) representation of the map is that it is not convenient for crosstalk characterization. There is no direct meaning in the values of the different terms. For example, a completely separably map, i.e. $\eta=0$, will have non-zero $\alpha$ terms for $|S|>1$ because $\alpha_S = \left(\prod_{\{i\subseteq B_m||i|=1\}} \alpha_{i}\right)^{|S|/m}$ for $|S|>1$ (e.g. $\alpha_{111}=(\alpha_{100}\alpha_{010}\alpha_{001})^{1/3}$). Therefore, to characterize the weight and locality of the twirled noise channel $\tilde{\Lambda}$, we attempt to find an alternative parameterization. The parameterization we introduce factors out independent error contributions of successively higher weights. We refer to 
\begin{eqnarray}
	\Lambda_S(\rho) & = & (1-\epsilon_S)\rho + \frac{\epsilon_S}{\prod_{\{i|i\in S\}} (4^{n_i}-1)) +1 }\left(\rho + \calw_S(\rho)\right) \nonumber \\
	 & = & (1-\epsilon_S)\rho + \frac{\epsilon_S}{m_S}\left(\rho + \calw_S(\rho)\right) \label{eqn:fixedweightmap}
\end{eqnarray}
as the {\it fixed-subspace-weight depolarizing channel}. This channel is CPTP if and only if $\epsilon_S\in [0,m_S/(m_S-1)]$. Defined in this way, the weight-1 subspace map is a fully depolarizing map, i.e., $\Lambda = (1-\epsilon) \rho + \epsilon/d \mathbf{I}$ in the subspace. We refer to a channel as a {\it weight-parameterized channel} if it has the form
\begin{equation}
	\Lambda_{\{\epsilon_S\}}=\underset{T\subseteq B_m}{\bigcirc}\Lambda_T, \label{eqn:weightmap}
\end{equation}
i.e. if it is the composition of fixed-subspace-weight error channels. The weight-parameterized channel is well-defined since the \PTM of the fixed-weight error channels are diagonal and therefore mutually commute. The channel has $2^m-1$ parameters $\{\epsilon_S\ |\ S\subseteq B_m, S\neq\emptyset\}$ since $\Lambda_\emptyset(\rho)=\rho$ for all choices of $\epsilon_\emptyset$. Every weight parameterized map of the form (\ref{eqn:weightmap}) is also a twirled map and $\{\alpha\}$ are related to $\{\epsilon\}$ as,
\begin{equation}
	\alpha_S = \prod_{T\subseteq [m], S\cap T\neq\emptyset} (1+y_S(T)\epsilon_T), \label{eqn:alpha_eps}
\end{equation}
where $y_S(T)$ is a set of factors that depend only on the way the subspace is divided. This can be seen by converting 
(\ref{eqn:fixedweightmap}) into the \PTM form,
\begin{equation}
	(R_{\Lambda_S})_{i,i} = (1-\epsilon_S) + \epsilon_S (R_{\mathrm{depol},S})_{i,i}
\end{equation}
where 
\begin{eqnarray}
	(R_{\mathrm{depol},S})_{i,i} =  \frac{1}{\prod_{\{i|i\in S\}(4^{n_i}-1)) +1 }} \nonumber \\
	\left(1 + \sum_{\substack{\PP_j\in\calP_n\\\mathrm{supp}(\sigma)=S}} (-1)^{\omega(\PP_i,\PP_j)}\right).
\end{eqnarray}
In the \PTM representation compositions are multiplications, so (\ref{eqn:weightmap}) is
\begin{eqnarray}
	(R_{\Lambda_{\{\epsilon_S\}}})_{i,i} & = &  \prod_{S} \left[(1-\epsilon_S) + \epsilon_S (R_{\mathrm{depol},S})_{i,i}\right] \\
																 & = & \prod_{S}\left[ 1 + \epsilon_S \left((R_{\mathrm{depol},S})_{i,i}-1\right) \right]
\end{eqnarray}
and is precisely the form of (\ref{eqn:alpha_eps}). Given a set of measured $\{\alpha\}$ from \crb, we try and find the set of $\{\epsilon\}$ by inverting (\ref{eqn:alpha_eps}). In Fig.~\ref{fig:simdata} we show the set of $\{\epsilon\}$ for simulated data. \\

\textbf{Summary of the Correlated Randomized Benchmarking procedure} for an $n$-qubit system partitioned into $m$ subsystems ($B_m = \{\{i,j,\ldots\},\ldots,\{k,l,\ldots\}\}$). 
\begin{enumerate}
\item For each subsystem select $l-1$ random $n_i$-qubit Clifford gates. Calculate the inverse gate.
\item Apply each sequence of $l$ random Cliffords in parallel to the $n$-qubit system. 
\item Measure the $2^m$ Pauli-$\mathbf{Z}$ correlators as a function of $l$ for all subsystems $S\subseteq B_m$.
\item Repeat for different random sequences and average. Fit the correlator data to $A\alpha_S^l+B$ to get the set of $\{\alpha_S\}$.
\item  From the set of measured $\{\alpha_S\}$, find the set of $\{\epsilon_S\}$ by inverting (\ref{eqn:alpha_eps})
\end{enumerate}

In certain cases, there are twirled channels that cannot be represented as weight-parameterized channels. For example, the channel that results from twirling
\begin{equation}
{\cal K}(\rho) = \left(1-\frac{p+q}{2}\right)\rho + \frac{p}{2}X_1\rho X_1 + \frac{q}{2}X_2\rho X_2
\end{equation}
cannot have a physical $\epsilon_{\{1,2\}}$ parameter for any $p, q\in (0,1)$. This is easy to see. Let ${\cal K}_j(\rho) = (1-\kappa_j)\rho + \kappa_j X_j\rho X_j$, $j\in \{1,2\}$ and ${\cal K}_{1,2}(\rho) = (1-\kappa_{1,2})\rho+\kappa_{1,2}X_1X_2\rho X_1X_2$. Then
\begin{eqnarray}
	& & {\cal K}_{1,2}(({\cal K}_1\otimes {\cal K}_2)(\rho)) = \nonumber \\
& &\left[ (1-\kappa_1)(1-\kappa_2)(1-\kappa_{1,2})+\kappa_1\kappa_2\kappa_{1,2}\right]\rho \nonumber \\
& & + \left[\kappa_1(1-\kappa_2)(1-\kappa_{1,2})+\kappa_2(1-\kappa_1)\kappa_{1,2}\right]X_1\rho X_1 \nonumber \\
& & + \left[\kappa_2(1-\kappa_1)(1-\kappa_{1,2})+\kappa_1(1-\kappa_2)\kappa_{1,2}\right]X_2\rho X_2 \nonumber \\
& & + \left[\kappa_1\kappa_2(1-\kappa_{1,2})+\kappa_{1,2}\right]X_1X_2\rho X_1X_2.
\end{eqnarray}
Looking at the weight-2 term, we must have $\kappa_1\kappa_2(1-\kappa_{1,2})+\kappa_{1,2}=0$ so $\kappa_{1,2}=\frac{\kappa_1\kappa_2}{\kappa_1\kappa_2-1}$. However, $\kappa_1, \kappa_2\in (0,1)$, so $\kappa_{1,2}<0$. This simple channel exhibits correlations that decay at a faster rate than independent noise. We leave to future work the question of characterizing the twirled channels that are not weight-parameterized channels. However for maps with standard experimental errors terms, e.g. $T_1, T_2$ and coherent Hamiltonian errors, the model is well-behaved. \\ 

\begin{figure}
\includegraphics[width=0.45\textwidth]{\figfolder{S1}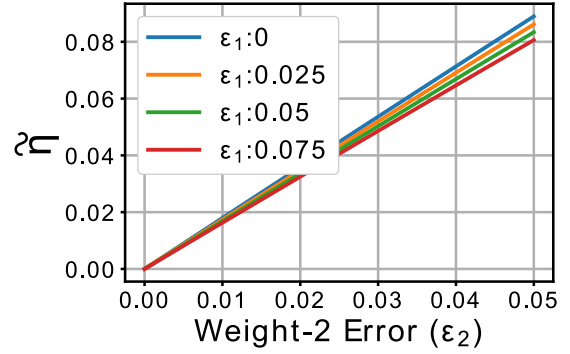}
\caption{(Color Online) Calculation of the crosstalk metric $\tilde{\eta}$ (\ref{eqn:eta2}) for a two-qubit weight-parameterized map versus the strength of the weight 2 errors for different values of the weight 1 errors. There is a weak dependence on the weight 1 errors. \label{fig:etadata}}
\end{figure}

The benefit of moving to the $\epsilon$ representation is that the locality and weight of the errors are clear. One question is whether the locality of the error coefficients is correctly reflected in the crosstalk metric $\eta$ (\ref{eqn:eta}), i.e., is the crosstalk metric independent of weight 1 errors? To answer this question we attempt to compute the $\eta$ of a twirled channel from $\{\{0\},\{1\}\}$ \crb assuming it has a weight-parameterization. We postulate that the optimal solution is a local weight-parameterized map in which case there is an analytic form,
\begin{eqnarray}
\eta(\tilde{\Lambda}) & = & \underset{q_1,\dots,q_n\in [0,1]}{\mathrm{min}}\sum_{\sigma\in\calP_n}\left| p_\sigma(\tilde{\Lambda})- \right. \nonumber \\
& & \left. \prod_{j\notin\textrm{supp}(\sigma)}(1-q_j)\prod_{j\in\textrm{supp}(\sigma)}\frac{q_j}{3}\right|
\end{eqnarray}
 Suppose now that $\tilde{\Lambda}$ has a weight parameterization $\{\epsilon_S\}$. It is clear that if $\epsilon_S=0$ for all $S$ with $|S|>1$, then $\tilde{\Lambda}$ is a 1-local channel and $\eta(\tilde{\Lambda})=0$. We can calculate $\eta(\tilde{\Lambda})$ for $n=2$ and find
\begin{widetext}
\begin{eqnarray}
\eta(\tilde{\Lambda}) & = & \underset{q_1,\dots,q_n\in [0,1]}{\mathrm{min}}  \left( \left| q_1q_2-\epsilon_1\epsilon_2+\epsilon_{12}\left(\frac{\epsilon_1}{3}+\frac{\epsilon_2}{3}+\frac{8}{9}\epsilon_1\epsilon_2-1\right)\right|\right. \nonumber \\
&. & + \left|q_2(1-q_1)-\epsilon_2(1-\epsilon_1)-\epsilon_{12}\left(\frac{\epsilon_1}{3}-\epsilon_2+\frac{8}{9}\epsilon_1\epsilon_2\right)\right| \nonumber \\
 & & + \left|q_1(1-q_2)-\epsilon_1(1-\epsilon_2)-\epsilon_{12}\left(\frac{\epsilon_2}{3}-\epsilon_1+\frac{8}{9}\epsilon_1\epsilon_2\right)\right| \nonumber \\
& &  + \left.\left|(1-q_1)(1-q_2)-(1-\epsilon_1)(1-\epsilon_2)+\epsilon_{12}\left((1-\epsilon_1)(1-\epsilon_2)-\frac{1}{9}\epsilon_1\epsilon_2\right)\right|\right).
\end{eqnarray}
\end{widetext}
If $\epsilon_{12}>0$ then numerical examples show that the minimum is not achieved at $(\epsilon_1,\epsilon_2)$ and is a function of $\epsilon_1$, $\epsilon_2$, and $\epsilon_{12}$. Roughly speaking, the nearest 1-local channel has a higher error rate than the 1-local terms in the weight parameterization due to the presence of the 2-local term. We plot some numerically calculated values for $\tilde{\eta}$ for a two-qubit system with increasing weight-2 errors and differing values of weight 1 errors in Fig.~\ref{fig:etadata}. We can see, as deduced from above, that there is some dependence on the weight-1 terms, but that dependence is weak so the metric does give a good sense of locality.\\

Since the twirled map is a Pauli channel, this motivates a modified definition of $\eta$ where we minimize over the closest tensor product of Pauli channels, 
\begin{equation}
	\tilde{\eta}(\Lambda) := \underset{\Gamma_j}{\mathrm{min}}\lVert\tilde{\Lambda}-\bigotimes_{j}\tilde{\Gamma}_j\rVert_\diamond \label{eqn:eta2}.
\end{equation}
Since $\tilde{\Gamma}$ is a Pauli channel there is an analytic formula for the diamond norm~\cite{magesan:2012},
\begin{eqnarray}
	\tilde{\eta}(\Lambda) & := & \underset{\gamma_{i}}{\mathrm{min}}\sum_{i} \left| \lambda_{i}-\gamma_{i} \right| \label{eqn:eta3}\\
	\tilde{\Lambda} & = & \sum_{i} \lambda_{i} \PP_{i} \rho \PP_{i} \\
	\tilde{\Gamma} & = & \sum_{i} \gamma_{i} \PP_{i} \rho \PP_{i}
\end{eqnarray}
such that the set $\{\lambda_i\}$ is so $\tilde{\Lambda}$ is of the form of Eqn~(\ref{eqn:twirledmap}) and $\{\gamma_{i}\}$ is so that $\tilde{\Gamma}$ is a set of local maps. Numerics from two-qubits are consistent with Eqn~(\ref{eqn:eta3}) being equal to Eqn~(\ref{eqn:eta}) when $\Lambda$ is a Pauli channel. Herein we will use $\tilde{\eta}$ for our experimental work because it is numerically tractable.

\section{Experiment}

\begin{figure}
\includegraphics[width=0.45\textwidth]{\figfolder{1}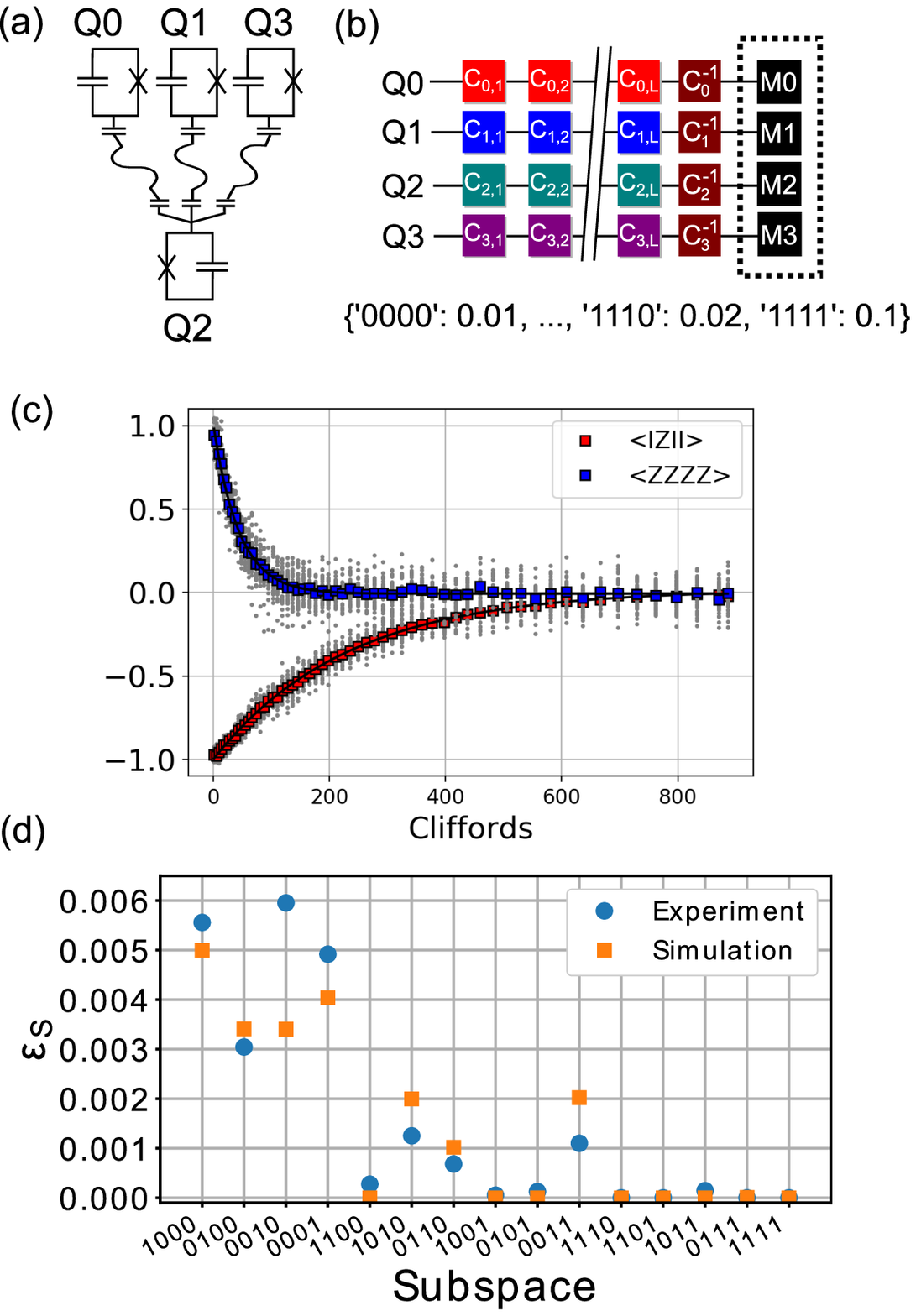}
\caption{(Color Online) (a) A schematic of the coupling between qubits on our four-qubit device. Q2 is coupled to the other qubits by separate coupling resonators. (b) A schematic of the $\{\{0\},\{1\},\{2\},\{3\}\}$ corrRB procedure. We perform  simRB and then correlate the measurement outcomes to produce a vector of probabilities in each of the $2^4$ states. (c) Representative decay data from the experiment for $\langle \mathbf{IZII} \rangle$ (what is measured in a standard RB experiment) and $\langle mathbf{ZZZZ}\rangle$. The data are fit to obtain the decay parameter $\alpha$. (d) The twirled map represented by the fixed-weight depolarization parameters $\epsilon$ (obtained from $\alpha$ using Eqn.~\ref{eqn:alpha_eps}). The simulation data from Fig.~\ref{fig:simdata} is shown for comparison.   \label{fig:schematic}}
\end{figure}

To test the corrRB method, we run $\{\{0\},\{1\},\{2\},\{3\}\}$ \srb (single qubit simRB) on a four qubit superconducting device. The qubits are fixed-frequency transmons~\cite{koch:2007} with frequencies $\omega = 2\pi \times \{5.367,5.243,5.185,5.357\}$~GHZ and typical coherences of $T_1=\{45,57,54,47\}~\mu$s and $T_2=\{74,100,91,81\}~\mu$s. As shown in (a) Fig.~\ref{fig:schematic}, qubits $0,1$ and $3$ are connected to qubit 2 by separate coupling buses. The qubits are coupled to these buses by a fixed interaction which results in an always-on \ZZ coupling term between the qubits~\cite{dicarlo:2009}. This term is measured to be $\{148,99,150\}$~kHz for qubits $0,1,3$ to qubit 2 and zero for qubits not on the same bus. As will be evident from the results, these \ZZ terms are the main source of crosstalk. Single qubit gates are implemented by applying DRAG shaped~\cite{gambetta:2011} microwave pulses to each qubit. Here we use the notation $X_{\pi/2}$ to refer to a $\pi/2$ rotation around the x-axis. Single-qubit gates on the different qubits are all the same length. To implement correlated RB (\crb) we perform the procedure outlined in the theory section, i.e., we perform simRB and measure the decay of the correlated polarization $\langle \prod_i (Z_i)^{q_i}\rangle$. For our particular experiments we measure this correlator by discriminating the state of each qubit into either a 0 or 1 after each run (``shot'') of the experiment, e.g. see Ref.~\cite{ryan:2015}. This is possible because each qubit has an independent readout cavity. We repeat and measure the experiment approximately 1000 times to build statistics and construct an estimate of the diagonal elements of a density matrix $\rho_{est}=\sum_{i} p_i |i\rangle \langle i|$ where $p_i=N_i/N$. The corrRB method is sensitive to correlations in the measurement and so, to remove these we apply a measurement correction matrix to our results. First, we measure $p_i$ after initalizing the system in each of the $2^n$ computational states. These vectors are estimates of the the rows of a matrix $\mathbf{A}$ which gives the probability of measureing a particular computational state for an arbitrary initial state $|\Psi\rangle$,
\begin{equation}
	\mathbf{p}_{meas} = \mathbf{A} \cdot \mathbf{p}_{|\Psi\rangle}
\end{equation}
where $\mathbf{p}_{|\Psi\rangle}$ is a vector of the diagonal elements of the density matrix. For perfect measurement $\mathbf{A}$ is the identity. The inverted $\mathbf{A}$ matrix ($\mathbf{A}^{-1}$) is the correction matrix, i.e.,
\begin{equation}
	\mathbf{p}_{state} = \mathbf{A}^{-1} \cdot \mathbf{p}_{meas},
\end{equation}
where $\mathbf{p}_{state}$ is the true probability vector. The correlation data presented in this paper is always corrected by this procedure. \\ 

We fit the measurement corrected $\mathbf{Z}$ correlators to the RB decay curve $A\alpha^{l}+B$ where $l$ is the number of Clifford gates in the RB sequence. In (c) Fig.~\ref{fig:schematic} we show experimental RB data for two such correlators for a $\pi/2$ pulse length of 96~ns. We fit each of the curves to obtain $\alpha_{exp}$ and then least squares minimize Eqn.~\ref{eqn:alpha_eps} against $\alpha_{exp}$ to obtain the fixed-weight depolarizing coefficients $\{\epsilon\}$ for the error map given by Eqn.~\ref{eqn:weightmap}. In (d) Fig.~\ref{fig:schematic} we plot these coefficients for the different subspaces, ordering by increasing subspace weight. In addition to the expected weight 1 errors (mostly due to T1 and T2) there are weight 2 terms, but essentially no weight 3 or 4 errors. For this data we calculate the crosstalk metric $\tilde{\eta}$ to be 0.0064. To compare to the experimental data we simulated these RB sequences by assuming perfect gate unitaries followed by a unitary for the \ZZ terms and a Kraus map for $T_1$ and $T_2$. The simulation data was discussed in the theory section and is shown in Fig.~\ref{fig:simdata}. We plot the fixed-weight depolarizing coefficients $\{\epsilon\}$ from the simulations with the values from the experiment in Fig.~\ref{fig:schematic} (d). There is good agreement between the results, particularly in the locality and weight of the higher order errors. This is compelling evidence that there are no additional crosstalk terms in the physical circuit and that the dominant source is the known \ZZ coupling. This demonstrates how correlated RB allows identification of crosstalk models. In a later section we will discuss how to reduce these crosstalk terms using an echo sequence and show that we can validate this reduction using the corr RB technique.

\section{Measuring Injected Noise}

\begin{figure}
\includegraphics[width=0.45\textwidth]{\figfolder{2}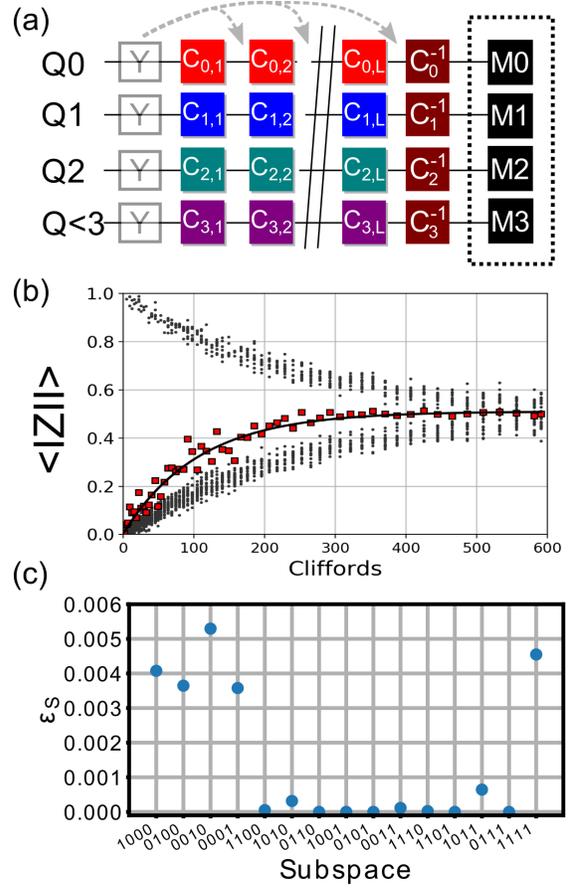}
\caption{(Color Online) (a) Schematic of the correlated RB sequence with injected weight 4 noise. For every trial (random iteration) of the RB sequence, a $X_{\pi}$ pulse (Y gate) on each qubit is probabilistically added after each Clifford which because a permanent part of the fixed sequence. (b) Typical RB curve with injected noise for the $\langle IZII\rangle$ correlator. Depending on the exact sequence of X gates added to the sequence the state should end up in $|1\rangle$ or $|0\rangle$. Averaging the results obtains a typical RB curve. (c) The fixed-weight depolarizing coefficients after fitting the data. The weight 4 depolarizing coefficient (0.0045) matches the probability of noise injection (0.005).  \label{fig:injectednoise}}
\end{figure}
 
To test the ability of the \crb technique to measure high weight errors we artifically inject crosstalk errors via probabilistic unitary gates. First, we generate a list of Cliffords corresponding to a random RB sequence. Then, after each Clifford in this sequence we add a $X_{\pi}$ (Y gate) on a subset of $m$ qubits with a probability $p$, i.e., a weight $m$ bit flip. This is shown schematically in (a) Fig.~\ref{fig:injectednoise}. We then run this new sequence on the device and measure the correlated outcomes. As per the RB protocol, we repeat this procedure for different random sequences and average. For each new trial we repeat the procedure to inject random bit flips and so each trial has a new random position of bit flips. Therefore, the randomization step of RB also randomizes the distribution of bit flips. A typical RB curve with injected noise is shown in Fig.~\ref{fig:injectednoise} (b). We see that individual RB runs jump between two curves depending on the culmutive action of the flips. On average, the curve decays exponentially. We then run the \crb analysis and plot the fixed-weight depolarizing coefficients $\{\epsilon\}$ in (c) Fig.~\ref{fig:injectednoise}. In this particular set of data the bit flips were injected on all four qubits with a probability of 0.005. The measured weight 4 depolarizing error is $0.0045$ matching well to the injection probability (for this data $\tilde{\eta}=0.011$). This shows that correlated RB is an effective technique for detecting up to the highest weight errors, and shows that we can reliably differentiate between errors of different weight. 

\section{Minimizing Weight 2 Errors with an Echo Sequence}

\begin{figure}
\includegraphics[width=0.45\textwidth]{\figfolder{3}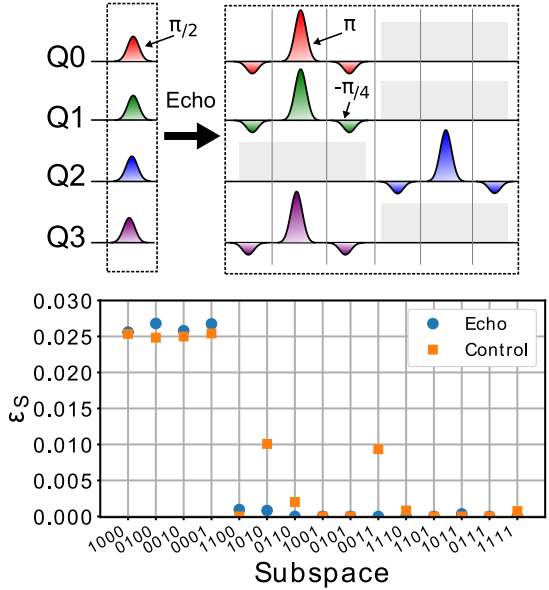}
\caption{(Color Online) (top) Schematic of the echo sequence. A concurrent set of $\pi/2$ pulses are split into a six-step sequence designed to echo the \ZZ terms for the circuit topology shown in Fig.~\ref{fig:schematic}. (bottom) Fixed-weight depolarizing coefficients (crosstalk errors) with the echo sequence (blue circles) and with a control sequence (orange squares). The value of the crosstalk metric $\tilde{\eta}$ the value decreased from 0.039 to 0.0037.  \label{fig:echo}}
\end{figure}

Given the ability to measure crosstalk errors with \crb, can the method be used to validate a crosstalk mitigation technique? Here we aim to reduce crosstalk using dynamic decoupling~\cite{lorenza:1999}, which minimizes coherent crosstalk errors using echo sequences. For example, to correct $Z$ errors on an idle qubit, we can apply a bit flip on the qubit at a regular intervals which flips the sign of $\langle Z \rangle$ so that on average it sums to zero. From the previous experiments we identified \ZZ errors are the dominant contributor to crosstalk errors. Therefore, we need to echo \ZZ terms, which is more complicated because we need to consider the collective qubit state. If the qubits were idle this would simply require applying a bit flip to Q2 at regular intervals (due to the connectivity as shown in Fig.~\ref{fig:schematic}). However, during an RB sequence each qubit is undergoing a random trajectory around the Bloch sphere thus making the echo sequence non-trivial. To avoid the complications of the random gates we idle certain qubits. Our sequence is as follows. For each set of four concurrent gates at time $t$ (our set of generating gates consists of $X/Y_{\pm \pi/2}$), we perform these gates in two times steps. In the first step, we do the gates on Q0,Q1 and Q3 while Q2 is idle. In the second time step we do the the Q2 gate with the other qubits are idle. To implement the echo, we replace each gate with an echoed version, 
\begin{equation}
	X/Y_{\pm \pi/2} = X/Y_{\mp \pi/4} \cdot X/Y_{\pm \pi} \cdot X/Y_{\mp \pi/4}. \label{eqn:echo}
\end{equation}
A schematic of this sequence is visualized in (a) Fig.~\ref{fig:echo}. The tradeoff is that the sequence is six times longer (if the pulses are kept the same length). Here we test the echo sequence experimentally with a pulse length of 59~ns (decoupling sequence length of 354~ns). First we generate simultaneous RB sequences and convert each of the pulses in the sequences to the echoed form using Eqn.~\ref{eqn:echo}. We then run the sequences on the experiment, analyze the correlated outcomes and obtain the fixed-weight depolarizing coefficients $\{\epsilon\}$. For comparison, we perform the same experiment without the echoed pulses, 
\begin{equation}
X/Y_{\pm \pi/2} = X/Y_{\pm \pi/4} \cdot X/Y_{0} \cdot X/Y_{\pm \pi/4},
\end{equation}
where $X/Y_{0}$ is an idle for the same duration as the normal gate. The coefficients from both experiments are plotted in (b) Fig.~\ref{fig:echo}. There is a clear reduction in the weight 2 crosstalk errors with the echo sequences. In terms of our crosstalk metric $\tilde{\eta}$ the value went from 0.039 to 0.0037. Compared to the data in Fig.~\ref{fig:schematic} the weight 1 error is higher because the effective pulse length is approximately 3.7 times longer. However, these weight 1 errors are easier to correct compared to weight 2 errors; this is captured by the reduction in the crosstalk metric. This experiment demonstrates how \crb can not only detect crosstalk errors, but can also validate mitigation techniques.

\section{Conclusions}

In conclusion, we have proposed and demonstrated a technique -- correlated RB -- that allows the identification of crosstalk errors in the average error map of particular weight and locality. The technique is simple to implement for any device that can already perform simultaneous RB as it is an extension that only requires correlating measurements and fitting to a specific error map. Here we demonstrated the technique for simultaneous RB on each qubit individually, but the method is general for simultaneous RB over any combination of subsystems and so should be able to measure crosstalk errors that occur during the two-qubit gates between subsystems (spectator errors). As presented here the number of correlators scales exponentially, $2^m$, but in practice not all correlators need to be included in the error model. \\

\begin{acknowledgments}
We thank Firat Solgun, Markus Brink, Sami Rosenblatt and George Keefe for modeling and fabricating the device. We thank Seth Merkel and Easwar Magesan for helpful discussions. This work was supported by the Army Research Office under contract W911NF-14-1-0124.
\end{acknowledgments}

\bibliography{corr_rb}

\end{document}